\documentclass[journal]{IEEEtran}
\usepackage[T1]{fontenc}
\usepackage[utf8]{inputenc}
\usepackage[english]{babel}
\usepackage{amsmath,amsfonts, amssymb}
\usepackage{mathtools}
\usepackage{acronym}
\usepackage{longtable}
\usepackage{textcomp}
\usepackage{multirow}
\usepackage{makecell}
\usepackage{subfigure}
\usepackage{graphicx}
\usepackage{color}
\usepackage{epsfig}
\usepackage{url}
\usepackage{balance}
\usepackage{float}
\usepackage{array}
\usepackage{verbatim}
\usepackage{pifont}
\usepackage{tabularx}
\usepackage{hyperref}
\usepackage{comment}
\usepackage{cite}
\usepackage{caption}

\usepackage[
	n,
	operators,
	advantage,
	sets,
	adversary,
	landau,
	probability,
	notions,	
	logic,
	ff,
	mm,
	primitives,
	events,
	complexity,
	asymptotics,
	keys]{cryptocode}

\newcommand{\cmark}{\ding{51}}
\newcommand{\xmark}{\ding{55}}

\newcolumntype{P}[1]{>{\centering\arraybackslash}p{#1}}

\acrodef{API}{Application Program Interface}
\acrodef{IC}{Integrated Circuit}
\acrodef{BLE}{Bluetooth Low Energy}
\acrodef{CA}{Certification Authority}
\acrodef{SDR}{Software Defined Radio}
\acrodef{DoS}{Denial of Service}
\acrodef{DH}{Diffie Hellman}
\acrodef{ECC}{Elliptic Curve Cryptography}
\acrodef{ECDH}{Elliptic Curve Diffie Hellman}
\acrodef{GNSS}{Global Navigation Satellite System}
\acrodef{GPS}{Global Positioning System}
\acrodef{IoT}{Internet of Things}
\acrodef{KDF}{Key Derivation Function}
\acrodef{HKDF}{HMAC Key Derivation Function}
\acrodef{MEO}{Medium Earth Orbit}
\acrodef{MITM}{Man-in-the-Middle}
\acrodef{PKC}{Public Key Cryptography}
\acrodef{PKI}{Public Key Infrastructure}
\acrodef{RF}{Radio Frequency}
\acrodef{RFID}{Radio Frequency Identification}
\acrodef{RSS}{Received Signal Strength}
\acrodef{SNR}{Signal-to-Noise-Ratio}
\acrodef{TLS}{Transport Layer Security}
\acrodef{PEPP-PT}{Pan-European Privacy-Preserving Proximity Tracing}
\acrodef{TCN}{Temporary Contact Numbers}
\acrodef{UHF}{Ultra High Frequency}
\acrodef{TDMA}{Time Division Multiple Access}
\acrodef{GFSK}{Gaussian Frequency Shift Keying}
\acrodef{DPSK}{Differential Phase Shift Keying}
\acrodef{FHSS}{Frequency-Hopping Spread Spectrum}
\acrodef{PACT}{Private Automated Contact Tracing}
\acrodef{RSS}{Received Signal Strength}
\acrodef{RSSI}{Received Signal Strength Indicator}
\acrodef{DP-3T}{Decentralized Privacy-Preserving Proximity Tracing}
\acrodef{LTE}{Long Term Evolution}

\graphicspath{{figs/}}

\title{Privacy-Preserving and Sustainable Contact Tracing Using Batteryless Bluetooth Low-Energy Beacons}
\author{
    \IEEEauthorblockN{Pietro Tedeschi\IEEEauthorrefmark{1}, Kang Eun Jeon\IEEEauthorrefmark{2}, James She\IEEEauthorrefmark{1}\IEEEauthorrefmark{2}, Simon Wong\IEEEauthorrefmark{2}, Spiridon Bakiras\IEEEauthorrefmark{1}, Roberto Di Pietro\IEEEauthorrefmark{1}}\\
    \IEEEauthorblockA{\IEEEauthorrefmark{1}Division of Information and Computing Technology, College of Science and Engineering, \\ Hamad Bin Khalifa University --- Doha, Qatar\\
    \IEEEauthorrefmark{2}HKUST-NIE Social Media Lab., The Hong Kong University of Science and Technology --- Hong Kong
    \\Email: \IEEEauthorrefmark{1}\{ptedeschi, sbakiras, pshe, rdipietro\}@hbku.edu.qa, \IEEEauthorrefmark{2}\{kejeon, eejames, tywongbf\}@ust.hk}\\
    }

\begin{document}

\maketitle

\captionsetup[figure]{name={Fig.}}

\begin{abstract}
Contact tracing is the techno-choice of reference to address the COVID-19 pandemic. Many of the current approaches have severe privacy and security issues and fail to offer a sustainable contact tracing infrastructure. We address these issues introducing an innovative, privacy-preserving, sustainable, and experimentally tested architecture that leverages batteryless BLE beacons.

\end{abstract}

\section{Introduction}
\label{sec:intro}
Smartphone-based contact tracing protocols~\cite{ahmed2020_access} have been adopted by many countries to help fight the spread of \mbox{\textsc{Covid-19}}. Most practical implementations today follow a common {\em modus operandi}: mobile devices continuously broadcast pseudo-random \ac{BLE} packets that are received and stored by other devices in the communication range; subsequently, the collected data are reconciled in either a centralized or decentralized fashion, in order to identify potential contagion events. 
The main challenge of contact tracing solutions is related to location tracking. Indeed, to signal the presence of a smart device, the devised solutions constantly spread pseudo-random packets via Bluetooth. Furthermore, every device maintains its own contact list by storing the signals broadcast by other devices.
However, this approach not only increases the energy burden on the user's smartphone---via constant \ac{BLE} scanning and broadcasting operations---but also inherently imperils user privacy.

Indeed, even though the user's packets are pseudo-random and change every few minutes, there is still a vulnerability window that allows an eavesdropping adversary to track the user's location. 
Such concerns are further amplified by incorrect software implementations, such as the Apple/Google privacy bug found in their
\textsc{Covid-19} exposure notification framework~\cite{apple_google_bug}. 
The above highlighted native privacy and energy concerns in existing solutions undermine the very purpose of contact tracing applications, hindering their adoption by the general public~\cite{cho2020contact}. 

To mitigate the aforementioned privacy and energy issues, we propose the deployment of a lightweight and wide-scale contact tracing infrastructure, consisting of \ac{BLE} transmitters. The packets transmitted by these devices would replace the smartphone-generated packets, but would still allow for accurate proximity tracing for the purpose of exposure notification. In particular, the users' smartphones would constantly intercept and store the infrastructure-based packets, thus gradually building a record of their precise location over time. Then, the exposure notification process would develop as in most standard decentralized BLE-based protocols. The benefits of our architecture are threefold: (i) unconditional privacy for users, since their devices are not emitting any information; (ii) reduced energy requirements for smartphones, which translates into longer battery life; and, (iii) potential for more accurate proximity detection, due to the presence of multiple (fixed) \ac{BLE} transmitters.

To facilitate an easy and wide-scale deployment, BLE beacons are typically battery-powered (similar to sensor network deployments). This latter point would trigger the issue of periodic battery replacement, which in turn would considerably increase the operational and maintenance cost of the infrastructure. 

Such overhead is further amplified in large-scale deployment cases. As an example, the Hong Kong International Airport had to deploy over $17,000$ beacons to provide indoor navigation services. By using \ac{BLE} beacons as a contact tracing infrastructure, energy consumption on user smartphones for broadcasting pseudo-random packets is off-loaded to the beacon infrastructure. More importantly, smartphones are not transmitting any information, so the users' privacy is unconditionally preserved---something hardly possible with existing device-to-device contact tracing protocols. Further, an infrastructure-based network can still support distributed protocols/frameworks of various contact-tracing solutions.

{\bf Contributions}.
We first show that energy-harvesting, batteryless beacons, are an affordable, reliable technology with respect to the operating cycle. We conducted an investigation on harvesting different types of energy sources, such as light, heat, and \ac{RF}, and also considered the corresponding energy harvesting architecture. Later, we embedded them within a comprehensive, viable architectural proposal to support contact tracing, and, finally, we showed experimental results supporting our findings. 
We also shed light on the trade-off between the broadcast frequency and transmit power that could affect the contact tracing performance and the batteryless beacon sustainability for a target tracing accuracy. We also provide a thorough discussion about the performance, efficiency, and security and privacy properties of our solution, while the paper concludes by highlighting future research directions.

\section{Related Work}
\label{sec:related}
In the following, we summarize the related work in the field, focusing on contact tracing approaches and energy harvesting technologies. We adopted the following terms throughout the paper:
\begin{itemize}
    \item \textit{BLE packet}: A broadcast packet sent using the BLE protocol.
    \item \textit{BLE beacon}: A piece of specialized hardware (not necessarily a smart device) that simply broadcasts BLE packets.
    \item \textit{luXbeacon}: A BLE beacon with energy harvesting capabilities to promote a self-sustainable operation.
\end{itemize}

\subsection{Digital Contact Tracing Solutions}
Nowadays, several governments, research institutes, and companies are working on exposure notification protocols to limit the spread of infectious diseases, such as \textsc{Covid-19}. Contact tracing is defined as an identification process that aims to track the recent physical contacts of individuals that have been tested positive for the virus. Broadly speaking, existing BLE-based contact tracing protocols can be categorized as follows.

\textbf{Decentralized Protocols}. In a decentralized architecture, users do not share any data with the authorities unless they have a confirmed positive test. In that case, the claimed positive device uploads its own transmitted BLE packets to the authorities' server. These packets are then propagated to the entire contact tracing network, where the individual smartphones perform the exposure notification function in a fully decentralized manner (by matching the published data against their own contact logs). Notable examples of decentralized contact tracing protocols are Apple/Google's framework~\cite{applegoogle} and the Decentralized Privacy-Preserving Proximity Tracing (DP-3T) protocol~\cite{troncoso2020}.

\textbf{Hybrid Protocols}. In a hybrid architecture, data collection follows the decentralized approach, i.e., each device maintains its private contact logs and does not disclose anything to the authorities. However, in hybrid protocols, the packets transmitted by the mobile devices are generated by the health authorities. Then, in the event of a positive test, the user's device discloses its contact logs to the authorities, and, therefore, exposure notification is performed by the authorities in a centralized manner. Typical examples of hybrid solutions are BlueTrace~\cite{bluetrace}---first adopted by Singapore---and the Pan-European Privacy-Preserving Proximity Tracing  (PEPP-PT) protocol~\cite{pepppt}.

\textbf{IoT-based Protocols}. IoT-based protocols employ an infrastructure of IoT devices to facilitate contact tracing. In other words, smartphones no longer interact with each other but rather depend on IoT devices to detect proximity. IoTrace~\cite{tedeschi2021_commag} is the only IoT-based solution to date. Under IoTrace, mobile devices are not required to scan the BLE channels for broadcast packets  sent by other devices. Instead, they simply broadcast their own packets, which are received and logged by the IoT infrastructure. The reconciliation mechanism is fully tunable and could range from a decentralized to a  centralized one. However, it is worth noting that reconciliation necessitates the transfer of a large number of packets to/from the centralized server, using 4G/\ac{LTE} communications. While the architecture introduced in this paper falls under the umberella of IoT-based solutions, its core functionalities, are very different from the ones provided by IoTrace.

\subsection{Energy-Harvesting Technologies for IoT Applications}
A BLE beacon can be configured with different advertising interval and transmit power values~\cite{Jeon2018_IoTJ}. The advertising interval determines the temporal spacing for the broadcast of beacon packets, while the transmit power controls its coverage area. A short advertising interval increases the beacon signal's reliability and enables more accurate distance estimation/localization. However, advertising intervals significantly influence the beacon's overall energy consumption and its lifetime.

In contact tracing applications, the energy demand for the devices is amplified due to various security and privacy requirements. For example, a static beacon may easily be spoofed or tracked, therefore, cryptographically secure hashing algorithms are often implemented on the device's firmware to periodically randomize the broadcast of  beacon packets~\cite{zidek2018beacon_security}. However, such an operation also leads to increased energy consumption and reduced lifetime.

\begin{figure}[ht]
	\centering
	\includegraphics[width = 0.5\columnwidth]{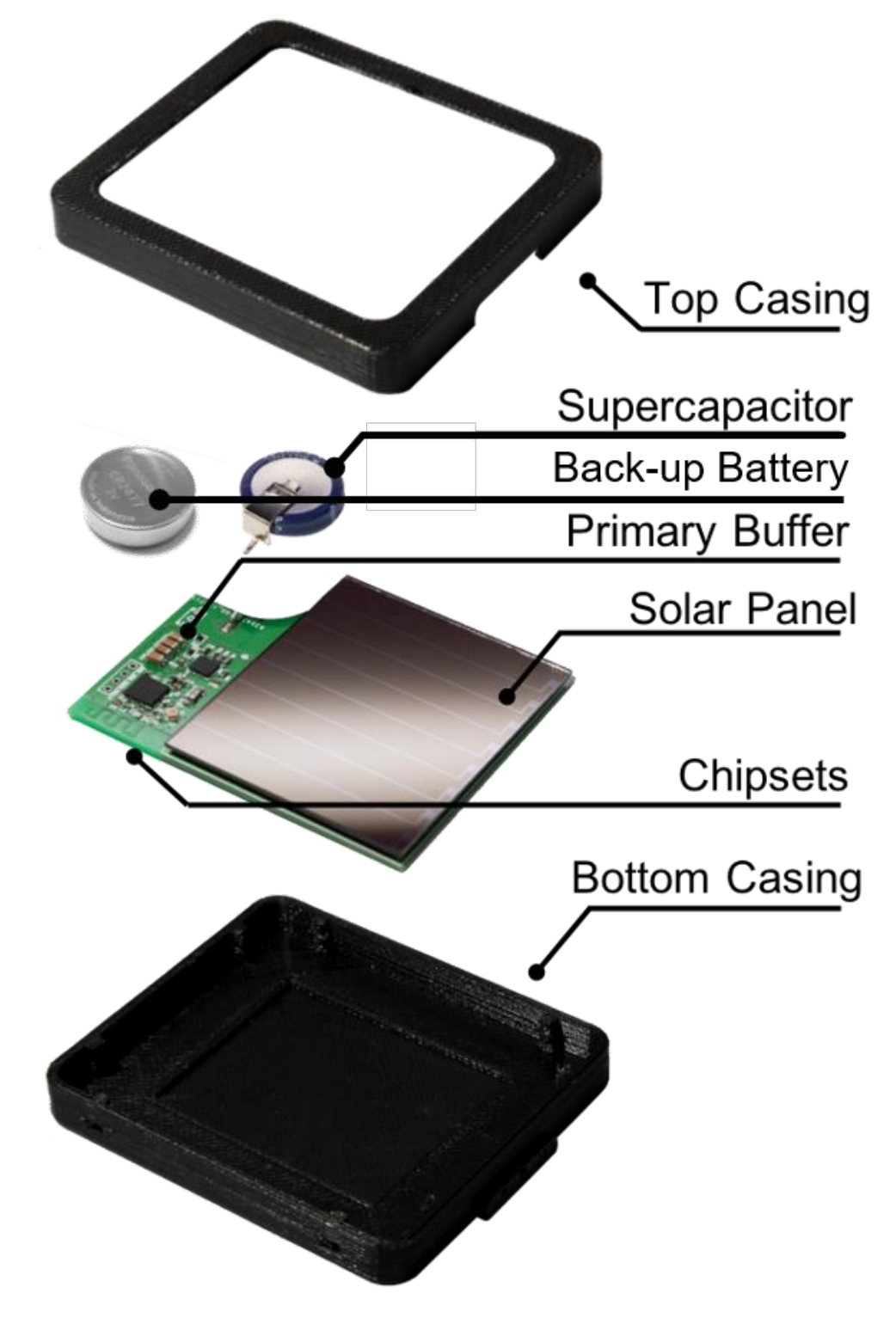}
	\caption{Circuit board and casing design of luXbeacon.}
	\label{fig:luxbeacon_board}
\end{figure}

To address these issues, we conducted an investigation on harvesting different types of energy sources, such as light, heat, and \ac{RF}, and also considered the corresponding energy harvesting architecture.
\textit{luXbeacon} is a \ac{BLE} beacon that can harvest and store ambient light energy for energy-neutral operation \cite{jeon2019luxbeacon}. It can operate in an indoor lighting environment with a minimum luminosity of $100$ lux, and is composed of $6$ major components, as shown in Fig.~\ref{fig:luxbeacon_board}:
\begin{enumerate}
    \item The solar panel harvests ambient light energy to power the load. The AM-1815 CA solar cell is optimized to harvest the visual light spectrum.
    \item Power management \ac{IC} routes the harvested energy from the solar panel to different parts of the circuit. The S6AE103A board leverages a linear harvesting architecture to achieve a low level of quiescent current (order of $nA$).
    \item The primary buffer is a small energy storage unit that is charged first with the harvested energy. The energy in the primary buffer is used to boot-up the Bluetooth \ac{IC}.
    \item The supercapacitor is a large energy storage unit, where the harvested energy is stored during an energy surplus. The stored energy is used to offset any energy deficit in the future.
    \item Bluetooth \ac{IC} is used to broadcast the \ac{BLE} beacon to the surrounding devices. 
    \item Back-up battery is used to power the luXbeacon when there is not enough ambient light energy to harvest and sustain its operation.
\end{enumerate}

\section{Threat Model}
\label{sec:threat}
In a BLE-based contact tracing application, the main threat to privacy is an eavesdropping adversary that collects all the transmitted packets. For instance, the adversary is equipped with either a \ac{SDR} with a powerful antenna, or a Bluetooth-compliant transceiver connected to a laptop/smartphone. Thus, the adversary only needs to set the frequency adopted by the Bluetooth communication technology to intercept all \ac{BLE} packets in the surrounding area. The attacker can also tag the packets with timestamp and geo-location information computed by standard GPS or indoor localization methods. An eavesdropping attack aims mostly at compromising the users' privacy by either tracking their movements or exposing their health status (with regards to the virus).

Alternatively, active adversaries may try to replay or relay previously transmitted packets to disrupt the operation of the contact tracing network. For example, the adversary may try to cause a large number of false-positive exposure notifications. Finally, we assume that the adversary can only perform polynomial-time computations and is unable to break the standard cryptographic primitives adopted in the pseudo-random packet generation functions.

\section{LuXbeacon Contact Tracing}
\label{sec:proposed_protocol}
The novelty of the proposed architecture lies in the deployment of a batteryless IoT infrastructure to facilitate privacy-preserving and energy-efficient proximity detection. In the following sections, we describe in detail the operations of the underlying contact tracing protocol.

\subsection{System Architecture}
\label{sec:sys_arch}
The entities involved in the proposed architecture are the following:

\emph{luXbeacon.} \ac{BLE}-based IoT device, equipped with specialized hardware for ambient-light energy harvesting. Every luXbeacon device broadcasts pseudo-random packets to the surrounding mobile devices.
    
\emph{User.} Smart device that runs the suggested contact tracing application. The app periodically scans the \ac{BLE} spectrum for packets transmitted by the deployed luXbeacon devices. Unlike existing approaches, the app operates in scan-only mode, i.e., it does not transmit any packets. During exposure notification, the smart devices approximate their relative proximity based on the received packets from the IoT infrastructure. 
    
\emph{Hospital.} Authorized medical facility that performs \textsc{Covid-19} infection tests. If a user tests positive, the health professionals are given permission to access their mobile device and forward the stored packets to the central authority.
    
\emph{Authority.} Trusted party whose role is to store the packets that were recently collected from the infected users. In a real scenario, this role can be played by the \emph{Ministry of Health}.

\subsection{Protocol Message Flow}
\begin{figure}[ht]
  \centering
  \includegraphics[angle=0, width=0.9\columnwidth]{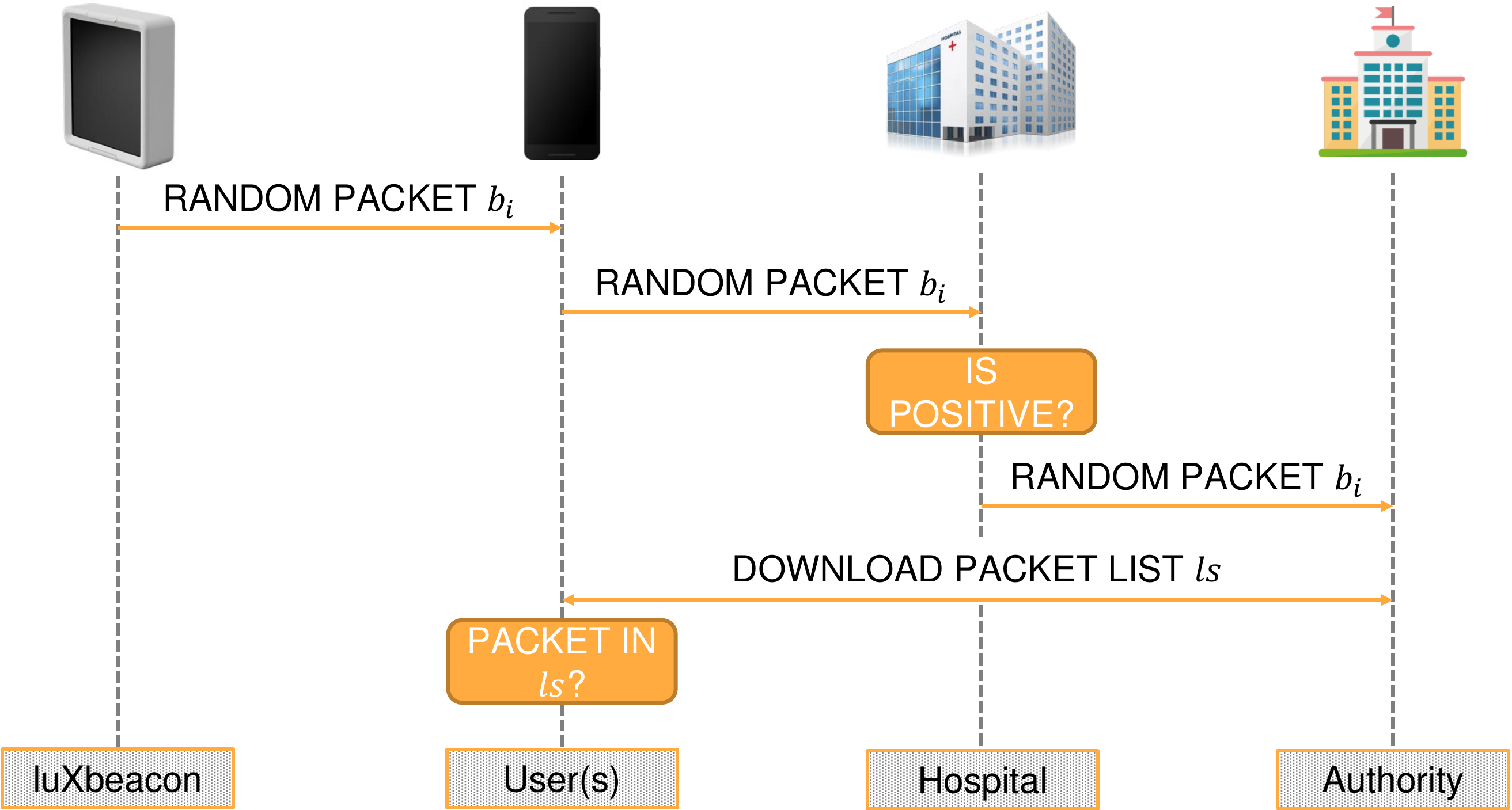}
  \caption{Message flow overview.}
  \label{fig:protocol}
\end{figure}

The protocol consists of two main tasks, namely, packet collection and exposure notification. We assume that each stored BLE packet at the user's device contains a timestamp, the luXbeacon's MAC address, a pseudo-random value (ephemeral ID), and the \ac{RSSI}. The high-level protocol message flow is as follows:
\begin{enumerate}
    \item Every \emph{luXbeacon} device periodically generates and transmits a pseudo-random \ac{BLE} packet, according to a secure keyed hash function.
    \item Every \emph{User} collects the packet(s) transmitted in its surrounding area. Should the \textit{User} test positive, the \textit{User} will send all its stored packets to the \emph{Authority}.
    \item Every \textit{User} periodically downloads the up-to-date packet list from the \emph{Authority}, and checks (locally) if there are common elements between its stored packets and the received list.
    \item Finally, for all identified common packets, the \textit{User} will estimate its relative proximity to that claimed positive, based on the signals' RSSI.
\end{enumerate}
The protocol message flow is also summarized in  Fig. \ref{fig:protocol}. Note that our architecture follows the decentralized exposure notification approach (Steps 3 and 4), where each device locally determines  whether the user was in close contact with a claimed positive. Assuming that the authorities will always learn the infected person's BLE packets, a further privacy goal is to \textit{not} disclose these packets to everyone else. Our discussion in Section~\ref{sec:sec_privacy} proposes such an approach that leverages well-known cryptographic protocols.

\subsection{Contact Detection and Result Notification} \label{sec:contact_detection_result}
In order to accurately detect a close contact between two users, it is critically important to estimate the following two parameters: (i) the distance between the two users; and (ii) the duration of the contact. The distance is essential because, if the two users were practising social distancing and separated by at least $2$--$3m$, the probability of contagion would be extremely low, and therefore, the contact would not be considered significant. Similarly, even if the two users were close enough for a contagion, but only for a period of less than a few minutes, the probability would also be very low. Therefore, the exposure notification function would consider these two variables when determining the threat level of a particular contact event. 

It is worth noting that both variables can be trivially estimated by the proposed architecture. First, the distance between two users can be approximated by observing and comparing the \ac{RSSI}s of their common packets. However, the \ac{RSSI} metric is subject to frequent fluctuations due to various environmental conditions, such as channel state, and fading and shadowing effects from the surrounding physical environment. Therefore, it is pivotal to deploy BLE beacons at a high density, in order to improve the distance estimation accuracy through better triangulation. On the other hand, the duration of contact can be acquired by simply computing (from the available timestamps) the time interval that encloses a certain subset of common packets.

\subsection{Comparison with Related Protocols}
\label{sec:related_protocols}
Table~\ref{tab:comparison} summarizes the characteristics of the most representative solutions reviewed in Section \ref{sec:related}---under different contact tracing architectures---and shows how they compare against the proposed protocol. First, our proposed architecture is the only one that replaces part of the smartphones' energetic cost (stemming from beacon transmissions) with renewable energy, hence having a low impact on the maintenance cost. This is not possible with IoTrace, because its energy demand for the IoT devices' operations is very high and cannot be supported by energy-harvesting technologies~\cite{tedeschi2020_comst}. For the same reason, IoTrace has a high maintenance/operation cost, due to the involvement of cellular communications and the need for frequent battery replacements.

\begin{table*}[htbp]
\caption{Comparison of state-of-the-art representative solutions. A \cmark\ symbol indicates the fulfillment of a particular feature, a \xmark\ symbol denotes that the feature is either not provided or not applicable.}
\centering
    \begin{tabular}{|c||c|c|c|c|}
    \hline
        \textbf{Features} & Decentralized protocols~\cite{applegoogle,troncoso2020} & Hybrid protocols~\cite{bluetrace,pepppt} & IoTrace~\cite{tedeschi2021_commag} & This work \\ \hline\hline
        \emph{Green Energy} & \xmark & \xmark & \xmark & \cmark \\ \hline
        \emph{Privacy Guarantee} & MEDIUM & LOW & MEDIUM & HIGH \\ \hline
        \emph{Total Energy Consumption} & $n\cdot(\alpha + \beta)$ & $n\cdot(\alpha + \beta)$ & $n\cdot\alpha + m\cdot(\beta + \gamma)$ & $n\cdot\beta + m\cdot\alpha$\\ \hline
        \emph{Maintenance/Operation Cost} & \xmark & \xmark & HIGH & LOW \\ \hline
    \end{tabular}
    \vspace{1ex}\\
    {\raggedright $\alpha$: \ac{RF} transmission cost, $\beta$: \ac{RF} receiving cost, $\gamma$: \ac{LTE} communication cost (with server), $n$: number of smartphones, $m$: number of IoT devices.}
\label{tab:comparison}
\end{table*}

In terms of privacy, hybrid protocols are the most vulnerable because the users' ephemeral IDs are generated by the central authorities, which results in low privacy guarantees. For example, a malicious adversary that compromises the centralized server is able to track the movements of all users. On the other hand, decentralized solutions (and IoTrace) are more privacy-preserving because users construct their own ephemeral IDs that are never revealed unless the user becomes infected with the virus. As such, they guarantee a medium level of privacy. Nevertheless, the broadcasting of packets from the mobile devices is, by itself, a privacy risk, as explained previously. On the contrary, our proposed architecture can guarantee a high level of privacy, as further discussed in Section~\ref{sec:sec_privacy}.

Finally, Table~\ref{tab:comparison} also shows a quantitative comparison of the energy consumption for the entire contact tracing architecture. Let $\alpha$ and $\beta$ be the daily \ac{RF} transmission and receiving costs (including channel scanning), respectively. Also, let $\gamma$ be the daily cost to communicate with the centralized server over an \ac{LTE} network. Then, the table shows the total daily energy consumption for a network with $n$ mobile devices and $m$ IoT devices. We expect that $\alpha \ll \beta \ll \gamma$, and $n > m$.

\section{Viability Study}
\label{sec:peva}
In this section, we revise the different requirements that assure the viability of a contact tracing protocol, showing that our approach satisfies them all.
\subsection{Sustainability}
\label{sec:sustainability}
\begin{figure}[t]
	\centering
	\includegraphics[width=0.9\columnwidth]{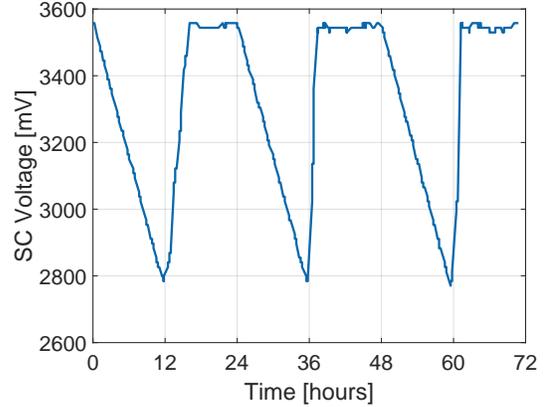}
	\caption{Supercapacitor voltage level of luXbeacon deployed in a real environment.}
	\label{fig:luXb_charge_cycle}
\end{figure}

The following section investigates and evaluates the energy efficiency and sustainability of luXbeacon, loaded with the contact tracing firmware---also performing the needed cryptographic operations. 
We first measured the power consumption of the contact tracing firmware, which proved to consume $12.2\mu A$, with $100ms$ advertising interval and $-8dBm$ transmit power. In order to prove its sustainability and practicality, we deployed a luXbeacon in a real-life environment and monitored the changes in its supercapacitor voltage. The luXbeacon was deployed near a window, to harvest both solar and indoor light sources, which can provide sufficient ambient power to support the luXbeacon. The result is shown in Fig.~ \ref{fig:luXb_charge_cycle}, where the luXbeacon  continuously charges and discharges its supercapacitor. It can also be observed that the supercapacitor voltage will never be lower than ~$2.7V$---the luXbeacon's operating voltage being $1.8V$. Such observation further supports the self-sustainability of the luXbeacon in a contact tracing application.

\begin{figure}[t]
	\centering
	\includegraphics[width=0.9\columnwidth]{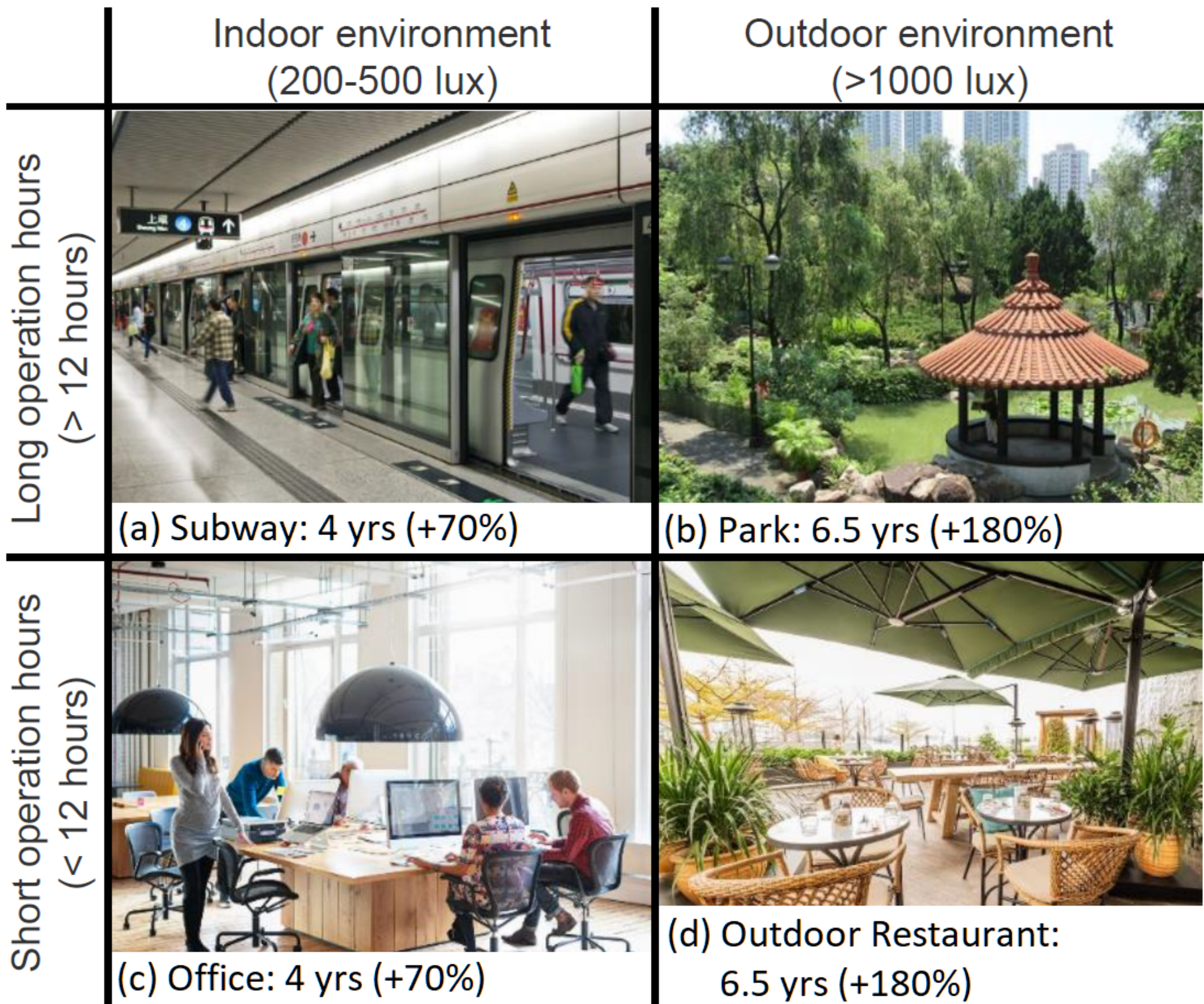}
	\caption{Expected lifetime of luXbeacon and lifetime extension compared to the battery powered devices, which last 2.3 years, under varying lighting conditions of social locations.}
	\label{fig:luXb_lifetime_table}
\end{figure}

To generalize our results, the lifetime of luXbeacon for various social locations was predicted using the lighting conditions of the locations. The predictions were made based on the measured energy consumption of the contact tracing firmware and also the power output of the solar panel. Fig.~\ref{fig:luXb_lifetime_table} shows $4$ different possible locations for deployment, with varying lighting conditions and operation hours. It can be seen that in all social locations, luXbeacon has an extended battery lifetime of at least $70\%$ compared to that of traditional battery-powered BLE beacons. 
Moreover, luXbeacon proved to be the most beneficial in outdoor deployment scenarios, which are the most difficult locations to conduct battery replacement or maintenance operations.

\subsection{Contact Tracing Accuracy}
\begin{figure}[htbp]
  \centering
  \includegraphics[angle=0, width=0.9\columnwidth]{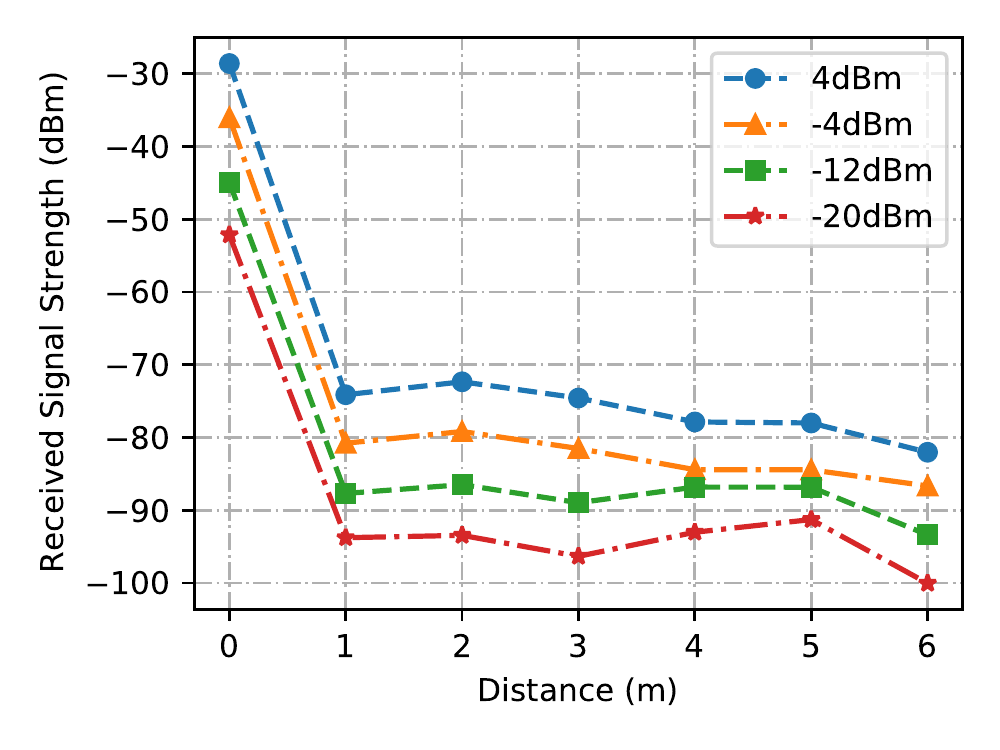}
  \caption{RSS of luXbeacon measured over distance.}
  \label{fig:rss_over_distance}
\end{figure}

In order to measure and reference the radio characteristics of luXbeacon, an experiment was conducted to investigate its \ac{RSS} over varying transmission power and receiving distance. The luXbeacon's \ac{RSS} was measured for $5$ minutes, for each distance ranging from $0m$ to $6m$ in $1m$ intervals. 
In Fig.~\ref{fig:rss_over_distance}, it can be observed that each transmit power curve is vertically separated from its neighbor by $5-7dBm$; however, the overall trend of the plot shows evident similarities. Also, it is noteworthy that the change in \ac{RSS} is dramatic between $0m$ to $1m$, but less noticeable after a $1m$ distance.

\begin{figure}[htbp]
  \centering
  \color{blue}
  \includegraphics[angle=0, width=0.9\columnwidth]{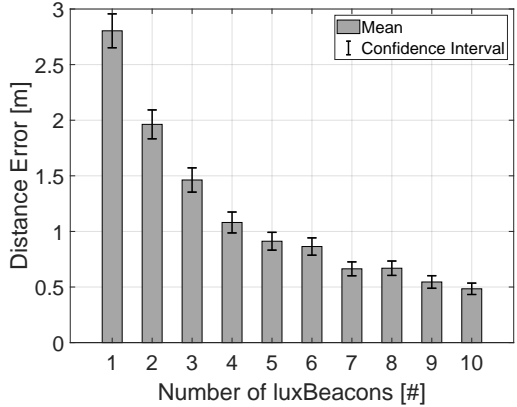}
  \caption{Error in distance estimation between two receivers.}
  \label{fig:accuracy}
\end{figure}

To validate our architecture, we also conducted an extensive simulation campaign using MATLAB\textcopyright 2020b, where we investigated how the random deployment of a varying number (from $1$ to $10$) of \ac{BLE} beacons could be leveraged for optimal coverage area and positioning accuracy. We present, for the first time in the literature, an end-to-end system that detects the contact between users based on BLE packet scanning information, namely \ac{RSSI} and ephemeral ID. Our architecture allows us to first estimate the distance of the users from the deployed BLE devices. From this information, our method then triangulates each user's position and estimates the distance between any two users with the distance error reported in Fig.~\ref{fig:accuracy}. The higher the number of deployed luxBeacons, the lower is the distance estimation error between two generic devices. Fig.~\ref{fig:accuracy} also reports the $95$\% confidence interval, computed over $10,000$ tests, with a luxBeacon TX power of $-8dBm$ and a random deployment of two smartphones in an area of $100m^2$. Let $R1$ and $R2$ be two generic receivers; the distance  between them, $\hat{d}$, is estimated as the Maximum Absolute Difference (MAD) between arrays $\mathbf{d_{R1}}$ and $\mathbf{d_{R2}}$, where each array consists of the estimated distances to each one of the surrounding luxBeacons, as shown in Eq.~(\ref{eq:accuracy}). The distances inside the two arrays are estimated by leveraging the relationship between \ac{RSSI} and distance (Fig.~\ref{fig:rss_over_distance}) collected from our experimental radio propagation model. Essentially, this approach is an attempt to estimate the distance between two users without knowing the precise locations of the surrounding luxBeacons.

\begin{equation}
    \label{eq:accuracy}
    \hat{d} = max(|\mathbf{d_{R1}} - \mathbf{d_{R2}}|)
\end{equation}

\subsection{Ephemeral ID Generation}
Each ephemeral ID is generated with the $\pcalgostyle{SHA256}$ hashing function and the XOR ($\oplus$) operation. The generated packet starts with the first $19$ bytes of device-specific information, such as device ID ($18$~bytes) and battery status information ($1$~byte). Then, the packet contains a timestamp of $8$ bytes. Further, we adopted the hashing function $\mathcal{H}$ on the concatenated $27$ bytes by providing an output of $32$~bytes. Finally, to reduce the size of the hashed data, we split the $32$ bytes of hashed data into two equals parts, and then we applied the $\oplus$ operation iteratively in order to reduce the hashed data to just $4$ bytes.

\section{Challenges and the Road Ahead}
\label{sec:challenges}
In the following sections, we describe the research challenges from the perspectives of security and privacy, infrastructure maintenance, and localization accuracy. We also outline the limitations of our proposed solution.

\subsection{Infrastructure and Maintenance Costs}
The proposed architecture requires a large deployment of IoT devices that are quite affordable when produced at mass-scale. As such, the main cost of the infrastructure will be determined by its maintenance. To this end, the adoption of energy-harvesting technologies, such as luXbeacon, reduces the maintenance cost significantly if deployed in an environment with sufficient light. However, in those environments that may not have enough light to enable energy-neutral operation, the energy consumption rate may vary due to non-uniform lighting condition, and so will the battery life. Such a phenomenon would lead to asynchronous expiry of battery lifetime, which may cause additional complications and difficulties in managing the infrastructure. To address such issues, it would be ideal to investigate and design energy-aware firmware that is capable of load-balancing to match its battery usage rate with that of nearby beacons.

A cost-benefit analysis for the proposed architecture should evaluate: (i) the efficiency of this new architecture in terms of resources; (ii) its effect on social well-being; and (iii) how social costs and benefits can be monetized.
When the luXbeacon is mass-produced, its unit cost will be  $\approx15.00\$$, including the casing and hardware, which is comparable to off-the-shelf battery-powered beacon devices that cost  $\approx30.00\$$~\cite{kontakt_price}.
It is also relevant to analyze the best deployment plan to cover the most crowded areas. Further, comparing our architecture to other \ac{BLE}-based approaches from the maintenance and application reliability perspective, the one-time cost to build the entire infrastructure can be considered relatively low.

\subsection{Tracing Performance}
\ac{BLE} beacon infrastructures have been widely used for various indoor localization applications. Many investigations have been performed on techniques that could enhance the positioning accuracy of a user in an environment with densely deployed IoT devices. However, very few studies exist concerning the energy consumption of a \ac{BLE} beacon. Since the luXbeacon's broadcasting frequency (i.e., advertising interval) is limited by the availability of harvestable ambient energy, the contact tracing accuracy may be affected by the scarce energy resources and the deployment environment. It would be imperative to study the relationship between luXbeacon's operational configurations---namely advertising interval and transmit power---with accuracy. Furthermore, the deployment method of the luXbeacon infrastructure may further be explored for optimal coverage area and positioning accuracy. Additionally, a method to accurately detect significant contacts between users must be investigated (e.g., user mobility). As future work, an evaluation of the luXbeacon's advertising interval and transmit power (i.e., the coverage area)---correlated to the density of a particular zone---is needed to achieve better performance in terms of energy consumption, communication efficiency, and hardware sustainability. This analysis allows for an implementation of a self-adaptive solution that permits tuning of the advertising interval, taking into account the area density as well as the beacon key update frequency.

\subsection{Security \& Privacy}
\label{sec:sec_privacy}
From a privacy perspective, the architecture follows the privacy-by-design approach. Indeed, off-loading the packet broadcast operation to the fixed hardware infrastructure avoids the ``data leakage'' issue for users, because their mobile devices are not transmitting any information. Therefore, our architecture makes it infeasible for an eavesdropping adversary to track users. However, if a user has a positive \textsc{Covid-19} test, the authorities have to publish their stored BLE packets to a public database for the purpose of exposure notification. As such, the user's recent location history is disclosed to the entire network. To this end, it is important to consider cryptographic techniques in the exposure notification function. In particular, instead of publishing the user's packets, the server could engage in a two-party private-set intersection protocol~\cite{decristoforo_2010} with individual users. The protocol's output would reveal (to the user) the common ephemeral ID set, but nothing else. It is also imperative to perform an experimental study to assess the effectiveness and computational cost of exposure notification in this privacy-preserving setting.

Further, compared to IoTrace, our architecture provides better security for data at rest, because no user information is stored on the luXbeacon(s). However, a critical security challenge is to find and analyze the right countermeasures to mitigate replay attacks. Specifically, a malicious adversary may deploy rogue luXbeacon devices to manipulate the protocol's proximity detection module. To this end, we should investigate the feasibility of detecting counterfeit beacons at the centralized server by analyzing the packets submitted by a new claimed positive. The analysis would consider the timing information, the beacons' ephemeral IDs (which are generated based on secret luXbeacon IDs), and the locations of the luXbeacon(s) that are known to the authorities.

\subsection{Discussion} 
\label{sec:discussion}
Besides the privacy-preserving benefit of the proposed architecture, the energy consumption of user smartphones is also reduced compared to existing methods. This is because they only need to carry out Bluetooth scanning operations, instead of both scanning and broadcasting.
While \ac{BLE} is a low-energy system compared to traditional Bluetooth, it involves a reasonably power-intensive operation. Continuous scanning would negatively affect the smartphone's battery life, and therefore, degrade the user's experience or even force them to uninstall the contact tracing app. Therefore, it is extremely important to consider the energy consumption at the end devices when developing the contact tracing system.

The energy consumption of the Bluetooth scanning operation depends on many factors, such as the Bluetooth SoC, the hardware design, the scanning parameters, and the number of scannable Bluetooth devices in the vicinity. Based on the nRF51822 SoC, an active and continuous scanning operation consumes $40mW$, whereas the broadcasting operation consumes at most $600\mu W$. As reported in~\cite{carroll2010mobile_energy}, the power consumption of Bluetooth scanning is similar to that of Wi-Fi during web browsing. To address such issues, the latest smartphone hardware and operating systems have implemented several mechanisms to minimize energy consumption. 
We should note that the scanning operation is duty-cycled at the OS level to reduce excessive power consumption. Therefore, it would be important to design a mobile app considering the energy consumption through an intelligent framework that requires minimum scanning operation.
Additionally, while a longer BLE packet broadcast cycle favours sustainability, it negatively impacts the proximity detection accuracy. There is a need to balance this trade-off, while also maintaining a low luXbeacon TX power. Nowadays, \ac{BLE} is the {\it de facto} standard  for the most prominent contact tracing solutions in the literature. Further research on various communication technologies is needed, e.g., Ultra-wideband carriers may be used to increase proximity tracing accuracy, privacy, and reliability.

\section{Conclusion}
\label{sec:conclusion}
The human and economic impact of the \textsc{Covid-19} pandemic has shown the need for novel technological solutions to tackle similar events in the future. Digital contact tracing can play a vital role in limiting the spread of deadly viruses. However, its effectiveness depends on its adoption by a large majority of the general public. To this end, privacy and energy-efficiency are two important metrics that can motivate users to participate in the contact tracing network. Our work makes a significant contribution towards this goal, by proposing an energy-efficient and privacy-preserving architecture for contact tracing. The proposed architecture leverages a dense deployment of batteryless IoT devices that constantly broadcast BLE packets for the purpose of proximity detection. We have shown that batteryless IoT has a reliable operating cycle and proved that their deployment can help  improve detection accuracy. The proposed architectural design enjoys low maintenance cost, reduces energy consumption on the user side, greatly improves distance accuracy estimation, and provides privacy by design. Finally, we have summarized the most important research challenges and directions that need to be addressed by the academia and industry, towards the development of IoT based privacy-preserving and efficient contact tracing.

\section*{Acknowledgements}
The authors would like to thank the anonymous reviewers, that helped improving the quality of the paper. This publication was partially supported by awards NPRP-S-11-0109-180242 from the QNRF-Qatar National Research Fund, a member of The Qatar Foundation, and NATO Science for Peace and Security Programme - MYP G5828 project ``SeaSec: DronNets for Maritime Border and Port Security''.

\bibliographystyle{IEEEtran}
\bibliography{main}

\begin{thebibliography}{10}
\providecommand{\url}[1]{#1}
\csname url@samestyle\endcsname
\providecommand{\newblock}{\relax}
\providecommand{\bibinfo}[2]{#2}
\providecommand{\BIBentrySTDinterwordspacing}{\spaceskip=0pt\relax}
\providecommand{\BIBentryALTinterwordstretchfactor}{4}
\providecommand{\BIBentryALTinterwordspacing}{\spaceskip=\fontdimen2\font plus
\BIBentryALTinterwordstretchfactor\fontdimen3\font minus
  \fontdimen4\font\relax}
\providecommand{\BIBforeignlanguage}[2]{{%
\expandafter\ifx\csname l@#1\endcsname\relax
\typeout{** WARNING: IEEEtran.bst: No hyphenation pattern has been}%
\typeout{** loaded for the language `#1'. Using the pattern for}%
\typeout{** the default language instead.}%
\else
\language=\csname l@#1\endcsname
\fi
#2}}
\providecommand{\BIBdecl}{\relax}
\BIBdecl

\bibitem{ahmed2020_access}
N.~{Ahmed} \emph{et~al.}, ``{A Survey of COVID-19 Contact Tracing Apps},''
  \emph{{IEEE Access}}, vol.~8, pp. 134\,577--134\,601, 2020.

\bibitem{apple_google_bug}
{Andrew Hayward - Decrypt}, ``{Privacy Bug Found in Apple, Google COVID-Tracing
  Framework},''
  \url{https://decrypt.co/40765/privacy-bug-found-apple-google-covid-tracing-framework},
  2020, accessed: 2021-07-30.

\bibitem{cho2020contact}
H.~Cho \emph{et~al.}, ``{Contact Tracing Mobile Apps for COVID-19: Privacy
  Considerations and Related Trade-offs},'' 2020.

\bibitem{applegoogle}
\BIBentryALTinterwordspacing
{Apple Google}. (2020) {Privacy-Preserving Contact Tracing}. (Accessed:
  2021-03-07). [Online]. Available:
  \url{https://www.apple.com/covid19/contacttracing}
\BIBentrySTDinterwordspacing

\bibitem{troncoso2020}
``{Decentralized Privacy-Preserving Proximity Tracing: Overview of Data
  Protection and Security},''
  \url{https://github.com/DP-3T/documents/blob/master/DP3T White Paper.pdf},
  2020, (Accessed: 2021-03-07).

\bibitem{bluetrace}
J.~{Bay} \emph{et~al.}, ``{BlueTrace: A privacy-preserving protocol for
  community-driven contact tracing across borders},'' \emph{Government
  Technology Agency-Singapore, Tech. Rep}, 2020.

\bibitem{pepppt}
\BIBentryALTinterwordspacing
{PEPP-PT Team}. (2020) {Pan-European Privacy-Preserving Proximity Tracing}.
  (Accessed: 2021-03-07). [Online]. Available: \url{https://www.pepp-pt.org/}
\BIBentrySTDinterwordspacing

\bibitem{tedeschi2021_commag}
P.~Tedeschi \emph{et~al.}, ``{IoTrace: A Flexible, Efficient, and
  Privacy-Preserving IoT-enabled Architecture for Contact Tracing},''
  \emph{{IEEE Communications Magazine}}, 2021.

\bibitem{Jeon2018_IoTJ}
K.~E. {Jeon} \emph{et~al.}, ``{BLE Beacons for Internet of Things Applications:
  Survey, Challenges, and Opportunities},'' \emph{{IEEE Internet of Things
  Journal}}, vol.~5, no.~2, pp. 811--828, 2018.

\bibitem{zidek2018beacon_security}
A.~{Zidek} \emph{et~al.}, ``{Bellrock: Anonymous Proximity Beacons From
  Personal Devices},'' in \emph{2018 IEEE International Conf. on Pervasive
  Computing and Communications (PerCom)}, 2018, pp. 1--10.

\bibitem{jeon2019luxbeacon}
K.~E. {Jeon} \emph{et~al.}, ``{luXbeacon—A Batteryless Beacon for Green IoT:
  Design, Modeling, and Field Tests},'' \emph{IEEE Internet of Things Journal},
  vol.~6, no.~3, pp. 5001--5012, 2019.

\bibitem{tedeschi2020_comst}
P.~{Tedeschi} \emph{et~al.}, ``{Security in Energy Harvesting Networks: A
  Survey of Current Solutions and Research Challenges},'' \emph{{IEEE
  Communications Surveys Tutorials}}, pp. 1--1, 2020.

\bibitem{kontakt_price}
\BIBentryALTinterwordspacing
Kontakt.io. (2021) {Smart Beacon}. (Accessed: 2021-07-24). [Online]. Available:
  \url{https://store.kontakt.io/our-products/30-smart-beacon-sb16-2.html}
\BIBentrySTDinterwordspacing

\bibitem{decristoforo_2010}
E.~De~Cristofaro \emph{et~al.}, ``{Practical Private Set Intersection Protocols
  with Linear Complexity},'' in \emph{Financial Cryptography and Data
  Security}, R.~Sion, Ed.\hskip 1em plus 0.5em minus 0.4em\relax Berlin,
  Heidelberg: Springer Berlin Heidelberg, 2010, pp. 143--159.

\bibitem{carroll2010mobile_energy}
A.~Carroll \emph{et~al.}, ``{An Analysis of Power Consumption in a
  Smartphone},'' in \emph{Proceedings of the 2010 USENIX Conference on USENIX
  Annual Technical Conference}, ser. USENIXATC'10.\hskip 1em plus 0.5em minus
  0.4em\relax USA: {USENIX Association}, 2010, p.~21.

\end{thebibliography}
\section*{Biographies}
\noindent 
{\textbf{Pietro Tedeschi} is a PhD Student at HBKU-CSE-ICT, Doha-Qatar. He received his Master's degree with honors in Computer Engineering at Politecnico di Bari, Italy. His research interests cover security issues in UAVs, Wireless, IoT, and Cyber-Physical Systems.\\
\textbf{Kang Eun Jeon} is a PhD Student at  HKUST, Hong Kong. He received his B.Eng. degree in Electronic Engineering also at HKUST. His research interests include self-sustaining, secure, and social BLE beacons for IoT applications.\\
\textbf{James She} is the founding director of HKUST Social Media Lab., Hong Kong and an associate professor at HBKU-CSE, Doha-Qatar. His current research areas include Social and Multimedia Computing, Data Science and AI for Visual Creativity, IoT for Sustainable, Smart and Interactive Systems.
\\
\textbf{Simon Wong} is an M.Phil. student at HKUST, Hong Kong. His research interests include technologies on machine learning and signal processing for IoT devices. \\
\textbf{Spiridon Bakiras} is an associate professor of cybersecurity at HBKU-CSE, Doha-Qatar. His research interests include security and privacy, applied cryptography, and spatiotemporal databases. He has held teaching and research positions at Michigan Technological University, the City University of New York, the University of Hong Kong, and the Hong Kong University of Science and Technology.
\\
\textbf{Roberto Di Pietro}, ACM Distinguished Scientist, is a Full Professor of Cybersecurity at HBKU-CSE. His main research interests include Security and Privacy, Distributed Systems, Virtualization, and Applied Cryptography. In 2020 he received the Jean-Claude Laprie Award for having significantly influenced the theory and practice of Dependable Computing.
}
\end{document}